\documentclass[journal,onecolumn,12pt]{IEEEtran}
\usepackage{amsfonts,color,morefloats}
\usepackage{amssymb,amsmath,latexsym,amsthm}

\newtheorem{theorem}{Theorem}
\newtheorem{lemma}[theorem]{Lemma}

\newtheorem{definition}{Definition}

\begin{document}

\title{Determination of 2-Adic Complexity of Generalized Binary Sequences of Order 2\thanks{The work was supported by the National Natural Science Foundation of China (NSFC) under Grant 11701553.}}

\author{Minghui Yang\thanks{Minghui Yang is with the State Key Laboratory of Information Security, Institute of Information Engineering, Chinese Academy of Sciences, Beijing 100093, China, (Email: yangminghui6688@163.com).} and
Keqin Feng \thanks{Keqin Feng is with the department of Mathematical Sciences, Tsinghua University, Beijing, 100084, China, (Email: fengkq@mail.tsinghua.edu.cn).}
}

\date{\today}
\maketitle

\begin{abstract}
The generalized binary sequences of order 2 have been used to construct good binary cyclic codes \cite{D2}. The linear complexity of these sequences has been computed in \cite{D1}. The  autocorrelation values of such sequences have been determined in \cite{BA} and \cite{D}. Some lower bounds of 2-adic complexity for such sequences have been  presented in \cite{He} and \cite{S}. In this paper we determine the exact value of 2-adic complexity for such sequences. Particularly, we improve the lower bounds presented in  \cite{He} and \cite{S} and the condition for the 2-adic complexity reaching the maximum value.
\end{abstract}


\section{Introduction} \label{sec-intro}
~\\

Let $p$ and $q$ be two distinct odd primes, $Z_{pq}=\mathbb{Z}/ pq\mathbb{Z}$. The unit group of the ring $Z_{pq}$ is
\begin{eqnarray*}
Z_{pq}^{*}
&=&  \left\{ a \pmod {pq}: \gcd(a, pq)=1\right\}\\
&=& \left\{ ip+jq \pmod {pq}: 1\leq i \leq q-1, 1\leq j \leq p-1\right\}.
\end{eqnarray*}

Let $P=\{p, 2p, \ldots, (q-1)p\}$, $Q=\{q, 2q, \ldots, (p-1)q\}$. Then $Z_{pq}=Z_{pq}^{\ast}\bigcup P \bigcup Q \bigcup \{0\}$.
Let $S=\{s_i\}_{i=0}^{pq-1}$ be the binary sequence with period $pq$ defined by
\begin{eqnarray}\label{sequence}
s_i=\left\{ \begin{array}{ll}
0,                             & \mbox{ if $i\in Q\bigcup\{0\}$} \\
1, & \mbox{ if $i\in P$} \\
\frac{1}{2}\big(1-\big(\frac{i}{p}\big)\big(\frac{i}{q}\big)\big), & \mbox{ if $i\in Z_{pq}^{*}$ }
\end{array}
\right.
\end{eqnarray}
where $(\frac{\cdot}{\cdot})$ denotes the Legendre symbol.

 The linear complexity of these binary sequences has been computed in \cite{D1}, and the autocorrelation values of such sequences have been determined in \cite{BA} and \cite{D}. Such sequences have been used to construct good binary cyclic codes in \cite{D2}. Beside the linear complexity, the 2-adic complexity of a binary sequence is needed to be larger when the sequence is used as key in stream cipher of cryptographic system.

\begin{definition}\label{definition}
Let $S=\{s_i\}_{i=0}^{n-1}$ be a binary sequence with period $n$, $s_i\in \{0, 1\}$, $S(2)=\sum_{i=0}^{n-1}s_i2^i\in\mathbb{Z}$, and $d=\gcd(S(2), 2^n-1).$ The 2-adic complexity of $S$ is defined by
$$C_2(S)=\log_2\big(\frac{2^n-1}{d}\big).$$
\end{definition}

The 2-adic complexity of the sequences, defined by (1), is estimated in [5, 7] by presenting several lower bounds in general case, and determined in some particular cases. In this paper we determine the exact value of $C_2(S)$ in general case by proving the following result in the next section.

For a positive integer $m=2^am', 2\nmid m'$, we denote the odd part $m'$ of $m$ by $(m)_o$.

\begin{theorem}\label{theorem}
Let $p$ and $q$ be distinct odd primes, $S=\{s_i\}_{i=0}^{pq-1}$ be the binary sequence defined by the formula (\ref{sequence}) in Section \ref{sec-intro}. Then\\
(1) the 2-adic complexity of $S$ is
$$C_2(S)=\log_2\big(\frac{2^{pq}-1}{d_1d_2}\big)$$
where $d_1=\gcd((q-1)_o, 2^p-1)$ and $d_2=\gcd((p+1)_o, 2^q-1)$. Moreover, if $(q-1)_o<2p+1$ then $d_1=1$; if $(p+1)_o<2q+1$ then $d_2=1$.\\
(2) $C_2(S)\geq\log_2\big(\frac{2^{pq}-1}{\max ((q-1)_o, (p+1)_o)}\big).$ If both of $(q-1)_o<2p+1$ and $(p+1)_o<2q+1$ hold, then $C_2(S)$ reaches the maximum value $\log_2(2^{pq}-1).$
\end{theorem}
This result provides better lower bound on $C_2(S)$ than ones given in [5, 7], and improve the condition for $C_2(S)$ reaching the maximum values. Ref. \cite{He} used the method of autocorrelation function given in \cite{Hu}, and Ref. \cite{S} used the method of the determinant of a circulant matrix given in \cite{X} and Gauss periods. In the next section we present a formula on $S(2)\pmod {2^{pq}-1}$ directly from the definition formula (\ref{sequence}) of the binary sequence $S$ in terms of quadratic ``Gauss sums" $G_p$ and $G_q$ valued in $Z_m$ $(m=2^{pq}-1)$. It is possible to get the exact value of $C_2(S)$, and with much simpler computation.

\section{Proof of Theorem}\label{sec-proof}
~\\

Let $S=\{s_i\}_{i=0}^{pq-1}$ be the binary sequence defined by the formula (\ref{sequence}) in Section \ref{sec-intro}. Then
\begin{eqnarray*}
S(2)=\sum_{i=0}^{pq-1}s_i2^i
&\equiv & \sum_{i=1}^{q-1}2^{pi}+\frac{1}{2}\sum_{i=1}^{q-1}\sum_{j=1}^{p-1}\big(1-\big(\frac{ip+jq}{p}\big)\big(\frac{ip+jq}{q}\big)\big)2^{ip+jq}\pmod{2^{pq}-1}\\
&\equiv & \frac{2^{pq}-1}{2^p-1}-1+\frac{1}{2}(\sum_{i=1}^{q-1}2^{ip})(\sum_{j=1}^{p-1}2^{jq})-\frac{1}{2}\big(\sum_{i=1}^{q-1}\big(\frac{ip}{q}\big)2^{ip}\big)\big(\sum_{j=1}^{p-1}\big(\frac{jq}{p}\big)2^{jq}\big)\pmod{2^{pq}-1}.
\end{eqnarray*}
Let
$$G_q=\sum_{i=1}^{q-1}\big(\frac{ip}{q}\big)2^{ip}\in \mathbb{Z},\ \  G_p=\sum_{j=1}^{p-1}\big(\frac{jq}{p}\big)2^{jq}\in \mathbb{Z}.$$
Then
\begin{align}
2S(2) & \equiv2\cdot\frac{2^{pq}-1}{2^{p}-1}-2+\big(\frac{2^{pq}-1}{2^p-1}-1\big)\big(\frac{2^{pq}-1}{2^q-1}-1\big)-G_pG_q\pmod {2^{pq}-1}\notag\\
                                  &\equiv 2\cdot\frac{2^{pq}-1}{2^{p}-1}-2-\frac{2^{pq}-1}{2^p-1}-\frac{2^{pq}-1}{2^q-1}+1-G_pG_q\pmod {2^{pq}-1}\notag\\
                                 &\equiv \frac{(2^{pq}-1)(2^q-2^p)}{(2^{p}-1)(2^q-1)}-1-G_pG_q\pmod {2^{pq}-1} \label{equ2}
                                 \end{align}

\begin{lemma}\label{lemma1}
 (1) $G_p\equiv 0\pmod {2^q-1}$,  $G_q\equiv 0\pmod {2^p-1}.$\\
 (2) $G_p^2\equiv \big(\frac{-1}{p}\big)\big(p-\frac{2^{pq}-1}{2^q-1}\big)\pmod {2^{pq}-1}$, \  $G_q^2\equiv \big(\frac{-1}{q}\big)\big(q-\frac{2^{pq}-1}{2^p-1}\big)\pmod {2^{pq}-1}.$

\end{lemma}
~\\
\begin{proof}
(1) $G_q=\sum_{i=1}^{q-1}\big(\frac{ip}{q}\big)2^{ip}\equiv\sum_{i=1}^{q-1}\big(\frac{ip}{q}\big)=0\pmod {2^{p}-1}.$ Similarly we have $G_p\equiv0\pmod {2^{q}-1}.$

(2)\begin{eqnarray*}
G_p^2
&= & \sum_{i,j=1}^{p-1}\big(\frac{jq}{p}\big)\big(\frac{iq}{p}\big)2^{(i+j)q}\equiv\sum_{i,j=1}^{p-1}\big(\frac{ij}{p}\big)2^{(i+j)q}\pmod{2^{pq}-1}\\
&\equiv & \sum_{\lambda,j=1}^{p-1}\big(\frac{j^2\lambda}{p}\big)2^{(\lambda j+j)q}\equiv\sum_{\lambda=1}^{p-1}\big(\frac{\lambda}{p}\big)\sum_{j=1}^{p-1}2^{(\lambda+1) jq}\pmod{2^{pq}-1}\\
&\equiv & \big(\frac{-1}{p}\big)(p-1)+\sum_{\lambda=1}^{p-2}\big(\frac{\lambda}{p}\big)\sum_{j=1}^{p-1}2^{jq}\equiv\big(\frac{-1}{p}\big)(p-1)-\big(\frac{-1}{p}\big)\big(\frac{2^{pq}-1}{2^q-1}-1)\pmod{2^{pq}-1}\\
&\equiv & \big(\frac{-1}{p}\big)\big(p-\frac{2^{pq}-1}{2^q-1}\big)\pmod{2^{pq}-1}.
\end{eqnarray*}
Similarly we have $G_q^2\equiv \big(\frac{-1}{q}\big)\big(q-\frac{2^{pq}-1}{2^p-1}\big)\pmod {2^{pq}-1}.$
\end{proof}

Now we let
$$d=\gcd(S(2), 2^{pq}-1), d_1=\gcd(S(2), 2^{p}-1), d_2=\gcd(S(2), 2^{q}-1), d_3=\gcd(S(2), \frac{2^{pq}-1}{(2^p-1)(2^q-1)}).$$
Then we have $d|d_1d_2d_3$ and $d_i|d (i=1, 2, 3).$

~\\

\begin{lemma}\label{lemma2}(1) $\gcd(pq, d)=1.$ Namely, neither $p$ nor $q$ is prime divisor of $d$.\\
(2) $d_1, d_2$ and $d_3$ are coprime pairwisely. Therefore $d=d_1d_2d_3.$\\

\begin{proof}
(1) If $p|d$, then $p|S(2)$ and $p|2^{pq}-1$. From $2^p\equiv 2\pmod p$ we get $$0\equiv 2^{pq}-1\equiv 2^{q}-1\pmod p.$$ Then
$$\frac{(2^{pq}-1)(2^q-2^p)}{(2^p-1)(2^q-1)}\equiv \frac{(1-2^p)}{(2^p-1)}(1+2^q+2^{2q}+\cdots+2^{(p-1)q})\equiv -p\equiv 0\pmod p.$$
From the formula (\ref{equ2}) and Lemma \ref{lemma1} (\ref{sequence}) we have $0\equiv 2S(2)\equiv -1-G_pG_q\equiv -1\pmod p$. This contradiction shows that $p\nmid d.$ Similarly, if $q|d$, then $q|S(2), 2^q\equiv 2\pmod q$ and $$0\equiv 2^{pq}-1\equiv 2^p-1\pmod q. $$ Therefore
$$\frac{(2^{pq}-1)(2^q-2^p)}{(2^{p}-1)(2^q-1)}\equiv \frac{(2^q-1)}{(2^q-1)}(1+2^p+2^{2p}+\cdots+2^{(q-1)p})\equiv q\equiv0
\pmod q.$$
From the formula (\ref{equ2}) and Lemma \ref{lemma1}(1) we have $0\equiv2S(2)\equiv-1-G_pG_q\equiv-1\pmod q$. This contradiction shows that $q\nmid d.$
~\\

(2) From $\gcd(2^p-1, 2^q-1)=1$ we know that $\gcd(d_1, d_2)=1$. Suppose that $d_1$ and $d_3$ have a common prime divisor $\pi$. Then $\pi|2^p-1$ so that $\pi\nmid2^q-1$, and $\pi|\frac{2^{pq}-1}{(2^p-1)(2^q-1)}$. Therefore
$$0\equiv \frac{2^{pq}-1}{(2^p-1)(2^q-1)}=\frac{1}{2^q-1}(1+2^p+2^{2p}+\cdots+2^{(q-1)p})\equiv\frac{q}{2^q-1}\pmod \pi$$
which shows that $\pi=q$. This contradicts to $q\nmid d_1$ (Lemma \ref{lemma2} (\ref{sequence})). Thus we get $\gcd(d_1, d_3)=1.$  Similarly, suppose that $d_2$ and $d_3$ have a common prime divisor $\pi$. Then $\pi|2^q-1$,  $\pi\nmid2^p-1$ and
$$0\equiv \frac{2^{pq}-1}{(2^p-1)(2^q-1)}=\frac{1}{2^p-1}(1+2^q+2^{2q}+\cdots+2^{(p-1)q})\equiv\frac{p}{2^p-1}\pmod \pi$$
which shows that $\pi=p$. This contradicts to $p\nmid d_2$ (Lemma \ref{lemma2} (\ref{sequence})). Thus we get $\gcd(d_2, d_3)=1.$
\end{proof}
\end{lemma}

Now we determine $d_1, d_2$ and $d_3$. For a positive integer $m=2^lm', 2\nmid m'$, we denote the odd part $m'$ of $m$ by $(m)_o.$

\begin{lemma}\label{lemma3}(1) $d_1=\gcd((q-1)_o, 2^p-1)$. Moreover, if $(q-1)_o\leq2p-1$, then $d_1=1.$\\
(2) $d_2=\gcd((p+1)_o, 2^q-1)$. Moreover, if $(p+1)_o\leq2q-1$, then $d_2=1.$\\
(3) $d_3=1$.
~\\

\begin{proof} (1) By Lemma \ref{lemma1}(\ref{sequence}) we know that $G_pG_q\equiv 0\pmod {2^p-1}.$ Then from $\gcd(2^p-1, 2^q-1)=1$ and the formula (\ref{equ2}) we get
$$2S(2)\equiv \frac{(2^{pq}-1)(2^q-2^p)}{(2^p-1)(2^q-1)}-1\equiv (1+2^p+2^{2p}+\cdots+2^{(q-1)p})-1\equiv q-1\pmod {2^p-1}.$$
Therefore $d_1=\gcd(S(2), 2^p-1)=\gcd(q-1, 2^p-1)=\gcd((q-1)_o, 2^p-1)$.

Suppose that $d_1\neq 1$. Let $\pi$ be a prime divisor of $d_1.$ Then $\pi|(q-1)_o$ and $\pi|2^p-1$ so that the order of $2\pmod \pi$ is $p$. By the Euler Theorem we get $p|\pi-1$. Then we have $p\leq\frac{1}{2}(\pi-1)\leq \frac{1}{2}((q-1)_o-1)$. Namely, $(q-1)_o\geq 2p+1.$ Therefore if $(q-1)_o\leq 2p-1$ then $d_1=1.$
~\\

(2) By Lemma \ref{lemma1}(\ref{sequence}) we know that $G_pG_q\equiv 0\pmod {2^q-1}.$ Then from $\gcd(2^p-1, 2^q-1)=1$ and the formula (\ref{equ2}) we get
$$2S(2)\equiv \frac{(2^{pq}-1)(2^q-2^p)}{(2^p-1)(2^q-1)}-1\equiv \frac{1-2^p}{2^p-1}(1+2^q+2^{2q}+\cdots+2^{(p-1)q})-1\equiv -(p+1)\pmod {2^q-1}.$$
Therefore $d_2=\gcd (S(2), 2^q-1)=\gcd((p+1)_o, 2^q-1).$ Suppose that $d_2\neq 1.$ Let $\pi$ be a prime divisor of $d_2$. Then $\pi|(p+1)_o$ and $\pi|2^q-1$ so that the order of $2\pmod \pi$ is  $q$ and then $q|\pi-1$. Then we have $q\leq\frac{1}{2}(\pi-1)\leq\frac{1}{2}((p+1)_o-1).$ Namely, $(p+1)_o\geq 2q+1$. Therefore if $(p+1)_o\leq 2q-1$ then $d_2=1.$
~\\

(3) Suppose that $d_3$ has a prime divisor $\pi$. Then $\pi|S(2)$ and $\pi|\frac{2^{pq}-1}{(2^p-1)(2^q-1)}$. If $\pi|2^p-1$, then $\pi\nmid 2^q-1$ and
$$0\equiv\frac{2^{pq}-1}{(2^p-1)(2^q-1)}\equiv \frac{q}{2^q-1}\pmod \pi$$
which means that $\pi=q$. We get a contradiction to $q\nmid d_3$ (Lemma \ref{lemma2}(\ref{sequence})). Similarly, if $\pi|2^q-1$, we get $\pi=p$ which also contradicts to $p\nmid d_3$ (Lemma \ref{lemma2}(\ref{sequence})). Therefore the order of $2\pmod \pi$ is $pq$ and then  $pq|\pi-1$. On the other hand, from the formula (\ref{equ2}) we get $0\equiv S(2)\equiv -1-G_pG_q\pmod \pi$. By Lemma \ref{lemma1}(\ref{equ2}) we have $1\equiv G_p^2G_q^2\equiv\big(\frac{-1}{p}\big)\big(\frac{-1}{q}\big)pq\pmod \pi$. Namely, $\pi|pq-1$ or $\pi|pq+1$. Then we get $pq\leq\frac{1}{2}(\pi-1)\leq\frac{1}{2}(pq+1-1)=\frac{1}{2}pq,$ a contradiction. Therefore $d_3=1$. This completes the proof of Lemma \ref{lemma3}.
\end{proof}
\end{lemma}
~\\
Proof of Theorem \ref{theorem}
~\\

By Lemma \ref{lemma3} and $d= d_1d_2d_3$ (Lemma \ref{lemma2}(2)) we get $C_2(S)=\log_2\big(\frac{2^{pq}-1}{d_1d_2}\big)$ where $d_1=\gcd((q-1)_o, 2^p-1)$ and $d_2=\gcd((p+1)_o, 2^q-1)$. Moreover, both conditions of $(q-1)_o\geq 2p+1$ and $(p+1)_o\geq 2q+1$ can not hold simultaneously. By Lemma \ref{lemma3} we know that $d_1=1$ or $d_2=1$. Therefore $C_2(S)\geq\log_2\big(\frac{2^{pq}-1}{\max(d_1, d_2)}\big)\geq\log_2\big(\frac{2^{pq}-1}{\max((q-1)_o, (p+1)_o)}\big)$. Finally, if both of $(q-1)_o\leq 2p-1$ and $(p+1)_o\leq 2q-1$ hold, then $C_2(S)$ reaches the maximum value $\log_2(2^{pq}-1)$. $\hfill\blacksquare$

\end{document}